\RequirePackage{fix-cm}
\documentclass[smallextended]{svjour3}       
\smartqed  
\usepackage{graphicx}
\begin{document}

\title{Heavy Elements\thanks{All data available on request.}
}
\subtitle{They came out of the blue}

\titlerunning{Heavy Elements with CUBES}        

\author{Camilla Juul Hansen         \and
}


\institute{C. J. Hansen \at
Technical University Darmstadt, Institute for Nuclear physics,\\
              Schlossgartenstr. 2 (S2|11), 64289 Darmstadt \\
              \at
	      Goethe University Frankfurt, Institute for Applied Physics,\\
	      Max-von-Laue-Str. 12, 60438 Frankfurt am Main\\
              \email{hansen@iap.uni-frankfurt.de}    
}

\date{Received: 07 Oct 2021 / Accepted: 15 Feb 2022}

\maketitle

\begin{abstract}
How are the heavy elements formed? This has been a key open question in physics for decades. Recent direct detections of neutron star mergers and observations of evolved stars show signatures of chemical elements in the blue range of their spectra that bear witness of recent nuclear processes that led to heavy element production. 
The formation of heavy elements typically takes place through neutron-capture reactions creating radioactive isotopes, which following beta-decay turn into the stable isotopes we today can measure indirectly in the surfaces of cool, low-mass stars or meteoritic grains. The conditions (such as the neutron density or entropy) of these n-capture reactions remains to date poorly constrained, and only through a multidisciplinary effort can we, by combining and comparing observations, experiments, and theoretical predictions, improve on one of the top 10 most important open physics questions posed at the turn of the century. This emphasises the need for detailed observations of the near-UV to blue wavelength region. The shortage of spectrographs and hence spectra covering this range with high-resolution and high signal-to-noise has for decades played a limiting factor in our understanding of how heavy elements form in the nuclear reactions as well as how they behave in the stellar surfaces. With CUBES we can finally improve the observations, by covering the crucial blue range in more remote stars and also achieve a higher signal-to-noise ratio (SNR). This is much needed to detect and accurately deblend the absorption lines and in turn derive more accurate and precise abundances of the heavy elements. 
\end{abstract}

\section{Introduction}
\label{intro}
The vast majority of known elements are heavier than the Fe-group, yet much more is known about the elements lighter than Fe. To date, $\alpha-$element abundances of cool stars have been accurately and precisely measured in thousands of stars, while the heavy element information is much more limited and typically only known in hundreds or even tens of stars for some of these elements. Yet a common picture arises.
When comparing a heavy element like Ba to a lighter $\alpha-$element, e.g., Mg, a much larger star-to-star (Ba) abundance scatter is seen in spite of the limited sample sizes of stars with precise heavy element abundances. 
This scatter was hard to reconcile with the fact that elements like Mg, which are mainly produced in core collapse supernovae, showed a low scatter (e.g., \cite{Kobayashi2006}). Both elements (Mg and Ba) were thought to be produced in core collapse supernovae (ccSN) causing confusion when trying to model the abundances.  The abundance scatter is generally seen in Galactic chemical evolution studies focusing on heavy elements (e.g., \cite{Francois2007,Cescutti2008,Hansen2012,Roederer2013,Hansen2014b}). 

Improved nucleosynthesis, observationally derived abundances, and stochastic inhomogeneous GCE models showed multiple formation processes were likely needed to explain the behaviour of the heavy elements. Through detailed observations of up to 40 elements between Z=30 and 92 combined with improved yield predictions based on new and better reaction rates showed the existence of both weak and main neutron-capture processes through a slow ($s$) or rapid ($r$) capture channel -- or even an intermediate ($i$) neutron-rich environment. A few heavy isotopes are formed via proton captures (e.g., \cite{Pruet2006,Froehlich2006}).

Based on numerous detections of Sr and Ba (two of the best studied heavy elements owing to their relatively strong transitions in cool stars), it has been postulated that all metal-poor stars have been enriched in heavy elements (e.g., \cite{Roederer2014}) and that a cut-off or floor might exist (e.g, \cite{Aoki2013}). To date we do not know if this is an observational bias. Different populations of stars show very different Sr/Ba ratios (e.g., \cite{Francois2007}) and this ratio may serve as a useful tool to classify stars \cite{Hansen2019}. Other studies have shown that some dwarf galaxies may be missing a heavy element production channel (e.g., \cite{Koch2008,Koch2013,Mashonkina2017b}) while extreme Sr and Ba enrichment in stars belonging to some of the classical dwarf spheroidals is also found (e.g., \cite{Hansen2018}). A detailed heavy element inventory of low-mass stars in various populations and Galactic components is much needed to explore the behaviour of heavy element production through space and time.

\section{Observations of neutron-capture processes}
\begin{figure}
  \includegraphics[width=0.95\textwidth]{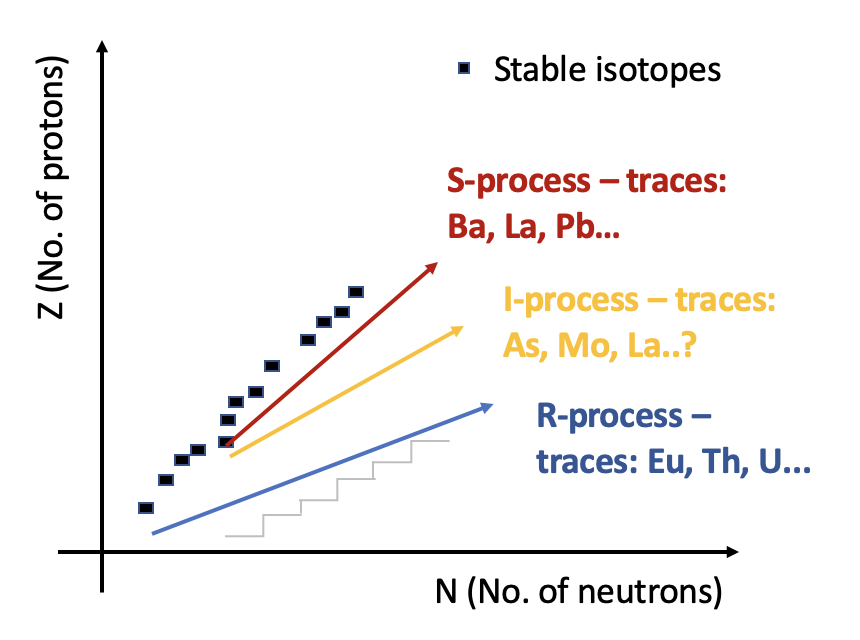}
\caption{Illustration of main neutron-capture channels (s, i, and r) and some of the chemical elements used to trace them. The stable isotopes are indicated as black squares, and the neutron drip line as a grey jagged line.}
\label{fig:Ncap}       
\end{figure}

In this section the focus lies with the heavy element (isotope) production taking place via neutron captures (see Fig.~\ref{fig:Ncap}).

\begin{figure}
  \includegraphics[width=0.95\textwidth]{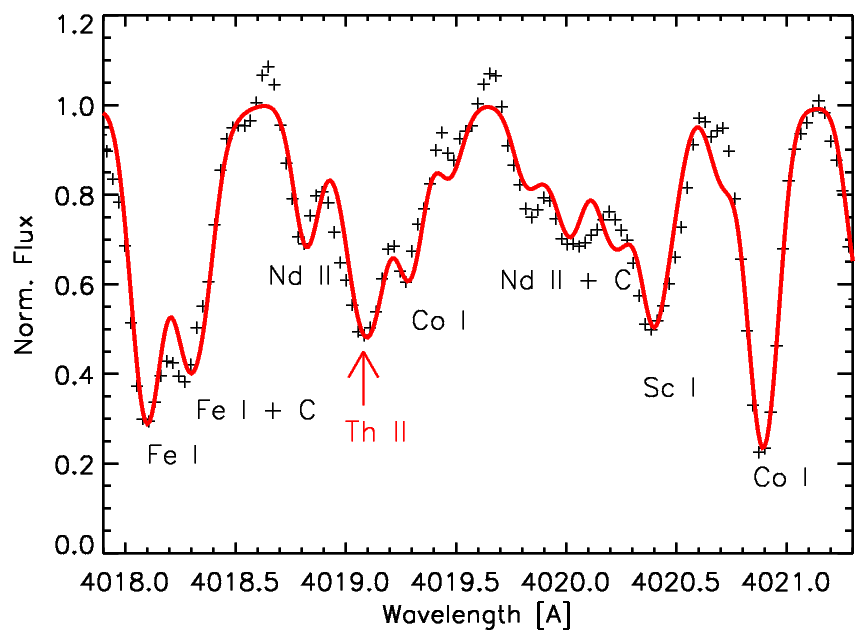}
\caption{High-resolution UVES spectrum of a metal-poor giant (+) in the Sagittarius dwarf Spheroidal showing the first detection of Th (red line) in such a galaxy. The region around 4000\,\AA\ is crowded and only the strongest lines are identified in the figure. }
\label{fig:Th}       
\end{figure}

\paragraph{The r-process} The observational studies of two extremely r-process enhanced, metal-poor stars \cite{Hill2002,Sneden2003} provided some of the first detections of the heaviest elements, like Th and U. These elements are to date only spectroscopically detected in tens of stars as they show weak lines just around 4000\,\AA\ (see Fig.~\ref{fig:Th}). The r-rich type of star is through its strong Eu enhancement often referred to as r-II stars \cite{Beers2005,Holmbeck2020}. The combination of Eu and Th has allowed for cosmochronologic age determinations \cite{Schatz2002,Cowan2002,Barbuy2011,Hansen2018}. Despite their sparsity, these stars have in the recent years been targeted in greater detail by smaller scale surveys such as RPA (the R-Process Alliance \cite{Hansen2018RPA,Sakari2018}) and new surveys like MINCE and CERES\footnote{Surveys emerged from the ChETEC network (www.chetec.eu) viz. "Measuring at Intermediate metallicity Neutron Capture Elements" (MINCE; PI G. Cescutti) and "Chemical Evolution of R-process Elements in Stars" (CERES; PI C. J. Hansen). }. However, a very different abundance pattern has been observed in other very metal-poor stars. Unlike the r-II stars, these show a, relatively speaking, excess of the lighter elements with respect to the heavy elements indicating the presence of an early n-capture process forming mainly lighter r-process elements \cite{Honda2004,Honda2007}. Observations of Sr, Y, Zr, Mo, Ru, Pd, and Ag solidified this finding showing the need for an additional process that contributed to the formation of weak r-process material early on in the Galactic history in metal-poor ([Fe/H]$<-2$) stars \cite{Francois2007,Hansen2012,Peterson2013,Hansen2014a,Peterson2020}. 
\paragraph{The s-process} The early detection of Tc in the blue spectral range in S-type stars was one of the first direct proofs of heavy element synthesis occurring inside asymtotic giant branch (AGB) stars. Technetium only has a very short-lived isotope and its presence in the stellar spectra bears witness of an ongoing s-process we never will detect directly otherwise \cite{Merrill1952,VanEck1999,Neyskens2015}. Various formation sites can host an s-process: the aforementioned AGB stars \cite{Busso1999,Kappeler2011} or fast rotating massive  stars (FRMS) \cite{Maeder2003,Meynet2006,Hirschi2007,Frischknecht2012}. However, the exact details of the underlying physics remains poorly constrained -- such as the impact of rotation or the size of the $^{13}$C-pocket in the AGB's He-intershell. Similar to the r-process, the s-process also has a weak channel, which may form elements as heavy as Mo or beyond \cite{Pignatari2013}. The exact extent of the elements formed through this channel is still not known. Moreover, the neutron-density or exposure in, e.g., AGB stars can be observationally tested through Pb abundances that are notoriously hard to derive in stellar spectra owing to the weak and blended Pb line at 4058\,\AA\ \cite{VanEck2001}. Detections of heavy elements like Mo or Pb will help us understand the s-process in greater detail.
\paragraph{The i-process} At an intermediate neutron exposure and density, the i-process \cite{CowanRose1977} runs between the s- and the r-process.  The process might take place in AGB stars following H-ingestion or in rapidly accreting white dwarfs \cite{Herwig2011,Hampel2016,Denissenkov2019}. This process is to date poorly known and details on its pattern are hard to extract from observations and theory.  I-process enhanced stars have been detected in various Galactic environments such as the halo \cite{Roederer2016}, in open clusters \cite{Mishenina2015}, and in the bulge \cite{Koch2019i}. Common to all of them is that various combinations of r- and s-process contributions cannot explain these chemically peculiar stars' abundance patterns. More complete stellar abundance patterns originating in various Galactic components will help clarify the nature of the i-process.

\section{Heavy elements in local stars and beyond}
\label{sec:1}
Not only the heavy element transitions reside in the blue wavelength range, also lighter elements like Be and numerous mono- and diatomic molecules show strong lines and bands in this region. Especially important are the N-strong bands, where NH only can be found in the blue. However, also C shows molecular bands in the blue. After H and He, O, C, and N are some of the most abundant elements in the universe. Detailed measurements of their molecules therefore serve as important (or unique) abundance indicators. Nitrogen is often missing from the stellar abundance pattern, because the blue range has not been observed. The ratio of C/N can also be used to trace stellar evolution and interior mixing processes (e.g. \cite{Spite2005}). Observing stars at lower and lower metallicities have shown that these are often Fe poor but C-N-O rich (e.g., \cite{Lee2013}) yet this is not always the case \cite{Caffau2011}.

\subsection{Carbon Enhanced Metal-Poor (CEMP) stars}
\label{sec:cemp}

The old, C-rich stars are called Carbon Enhanced Metal Poor (CEMP) stars, and they come in different subgroups depending on their exact C enhancement and heavy s- and/or r-process content. Originally, these stars were subclassified based on their C, Ba, and Eu abundances \cite{Beers2005}. However, detecting Eu in these stars with currently existing spectrographs and strong line blending is very challenging at low metallicity. Hence various sub-classifications have been suggested \cite{Masseron2010,Hansen2019}, where the latter makes use of strong-lined elements like Sr and Ba, which are considerably easier to detect. This also means that these trace elements can be observed in more remote parts of the Galaxy and possibly beyond. However, with an instrument like CUBES we are no longer restricted to a few elements; a richer chemical pattern of the heavy elements can finally be mapped with such an instrument. 

Understanding the formation of an element like Sr in detail is of great interest, as it is one of the lower-Z n-capture elements, which lies close to the first r- and s-peak, and must be by-passed by the formation path of heavier elements. Moreover, Sr can be formed in ccSN \cite{Winteler2012}, AGB stars \cite{Lugaro2012,Karakas2014}, massive fast rotating stars \cite{Meynet2006,Frischknecht2012}, and it was recently also detected directly in a neutron star merger \cite{Watson2019}. This was the first spectroscopic evidence of an ongoing r-process, thereby tying its occurrence to merger events. 
Barium or europium on the other hand will likely not be formed by the weaker processes and if so, only in small amounts. However, the extreme scatter in barium abundances (exceeding three orders of magnitude) is a sign of differing formation processes contributing at early times. The abundance scatter remains even after applying non-LTE corrections \cite{Korotin2015,Andrievsky2009} and seemingly also after applying 3D corrections \cite{Gallagher2020}. Hence, this scatter is not an observational artefact, and it is to date not understood. Most heavy elements have not been observed in great numbers with higher abundance precision to confirm this star-to-star abundance scatter. Moreover, the various CEMP subgroups only add to this star-to-star abundance scatter.

\subsection{Dwarf galaxies} 
Most of the CEMP stars or r-process enhanced stars are observed in the Milky Way halo. However, with the increasing number of detected dwarf Spheroidal (dSph) and ultra-faint dwarf (UFD) galaxies, and streams, r-rich stars have now also been observed and analysed in various kinds of environments (e.g., \cite{Aguado2021}). This has had a major impact on our understanding of galaxy formation and enrichment, as well as understanding the environment hosting the nuclear reactions. One of the strongest r-process enhancements was seen in the UFD Reticulum II \cite{Ji2016} which showed that an r-process event like neutron star mergers can also take place in small, dark-matter dominated systems and basically contribute the entire r-process inventory to the UFD. Observing the classical dwarfs in the Local Group often pushes UVES/VLT to its limits and requires long exposure times, thus the heaviest elements like Th have only been detected in Sagittarius \cite{Hansen2018} -- see also Fig.~\ref{fig:Th} -- while U has never been observed in a dwarf galaxy. Recently, the first detection of Lu in the dSph Fornax was made \cite{Reichert2021}. A higher efficiency and SNR, which CUBES will bring, is needed to map the n-capture elements in dwarf galaxies in order to understand how poorly studied elements like Pb and Hf or Lu  are formed in situ or ex situ. (Some of these elements have been detected in less than 10 stars \cite{Kobayashi2020}.)

\subsection{The metal-rich Galaxy}
At later times, that is in more metal-rich stars, for instance in the Galactic disk, it is challenging to explain the heavy element formation. Recent GCE models have explored the delay time of neutron star mergers and shown that one formation site alone cannot explain the Eu abundances derived in numerous disk stars \cite{Cote2019}. The time (stellar age) can in more metal-rich disk stars (solar twins) be traced via the [Y/Mg] ratio \cite{Spina2018} aiding our computation of stellar ages and improve on the age-metallicity degeneracy. Stellar abundances can also be compared to the chemical composition of comets (another CUBES science case) as well as the isotopic ratios in pre-solar grains, however, accurate and precise abundances are essential for this. Despite the Sun being the most observed star we know, several solar abundances remain uncertain and often the non-LTE and 3D corrections are not in agreement (one example is O \cite{Asplund2021,Caffau2015O,Bergemann2021}). To date about seven heavy elements can be fully corrected for the 1D LTE assumptions (e.g., \cite{Bergemann2012,Hansen2013,Gallagher2020}) but see also \cite{Hansen2020} and references therein.

\begin{figure}
  \includegraphics[width=0.95\textwidth]{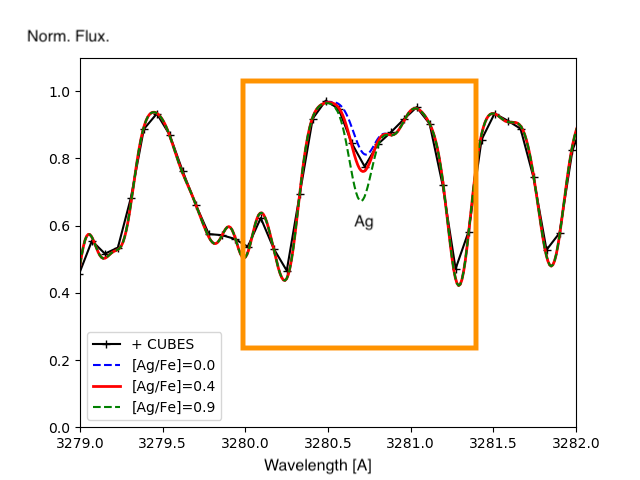}
\caption{Simulated CUBES spectrum (3\,\AA\,) around the Ag line including three different synthetic spectra with Ag varied as indicated in the legend. The orange contained region is compared to UVES spectra in Fig.~\ref{fig:Agspec}.}
\label{fig:blue}       
\end{figure}

\section{The blue wavelength range}
\begin{figure}
\hspace{-0.99cm}
  \includegraphics[width=1.2\textwidth]{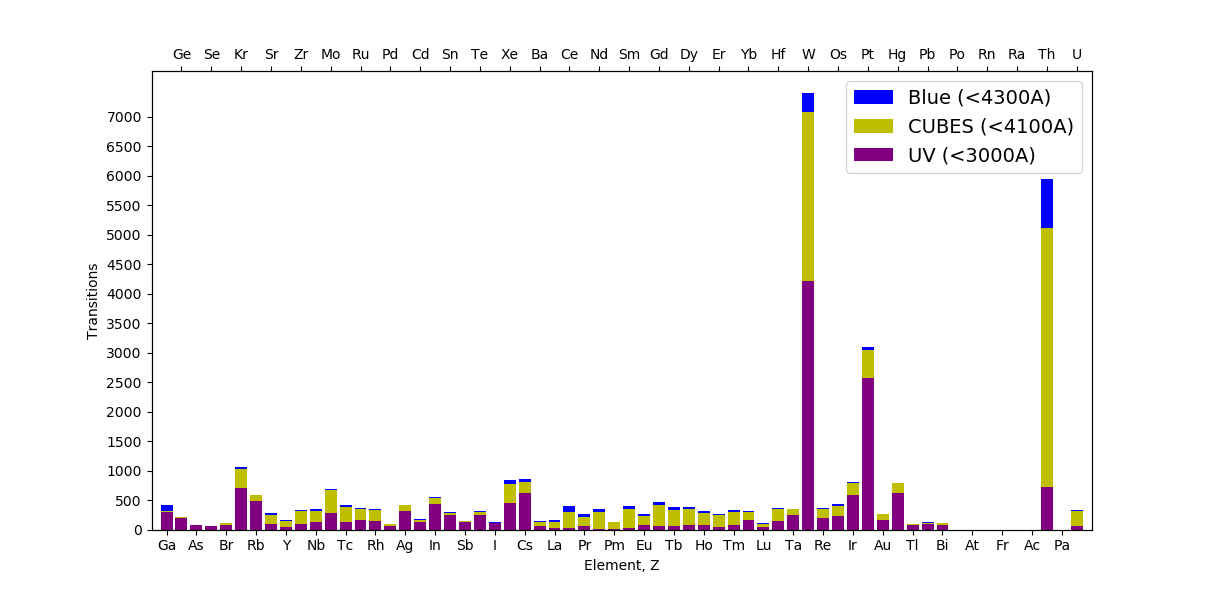}
\caption{The figure shows lines, which are available in the NIST atomic line database with some level of know atomic data. It reflects the distribution of atomic lines as a function of wavelength only, and not if the lines actually detectable in cool stars. However, the vast amount of atomic data available in the near-UV between 3000 and 4100\,\AA\, reflects that this range is most desirable to observe to obtain information on heavy elements. Lines detectable in cool, low-mass stars can be seen in Fig. 5 of \cite{Hansen2015} including also lines red-wards of 4300\,\AA\,. Their figure was created for 4MOST which will have a similar resolution to CUBES.}
\label{fig:hist}       
\end{figure}

The vast majority of the heavy elements (Z~$>30$) show their strongest transitions in the blue ($<4300$\,\AA\ \cite{Sneden2003} and see Fig.~\ref{fig:hist}) in cool stars where the density of absorption lines is very high, often causing line blending (for a 3\,\AA\ range in the blue see Fig.~\ref{fig:Th} and 3\,\AA\ in the near-UV see Fig.~\ref{fig:blue}). More details can be found in \cite{Hansen2015}.
A key to understanding the formation of heavy elements (at a nuclear level as well as in the grand scheme of Galactic enrichment) precise and accurate abundances are absolutely necessary. Here the resolving power, sampling,  and spectrum quality play important roles.
In \cite{Hansen2015} the abundance recovery of two blended (Gaussian shaped) lines was tested for 4MOST \cite{Caffau2013,Jong2014} which will have a resolution similar to CUBES. The tests were made generically and therefore expressed as a function of the full width at half maximum (FWHM) of the lines, which in turn reflects the resolution. The abundance of the two different lines could be perfectly recovered if the core-to-core separation was at least $0.6\times$FWHM, while a $\pm0.1$\,dex uncertainty could be recovered at $0.3\times$FWHM. If the abundance of one of the lines is constrained from other clean lines, this can be improved on to a full abundance recovery of the other line. Saturated lines, additional blends, or uncertain radial velocity shifts ($\sim1-2\,$km/s) would aggravate the case and could possibly lead to uncertainties as large as $\pm0.3-0.4$\,dex and $>1\times$FWHM is needed to properly separate the lines and derive precise abundances. Hence, a word of caution is in place when wanting to recover precise abundances in spectra that are of lower resolution and sampling.
\begin{figure}
  \includegraphics[width=0.9\textwidth]{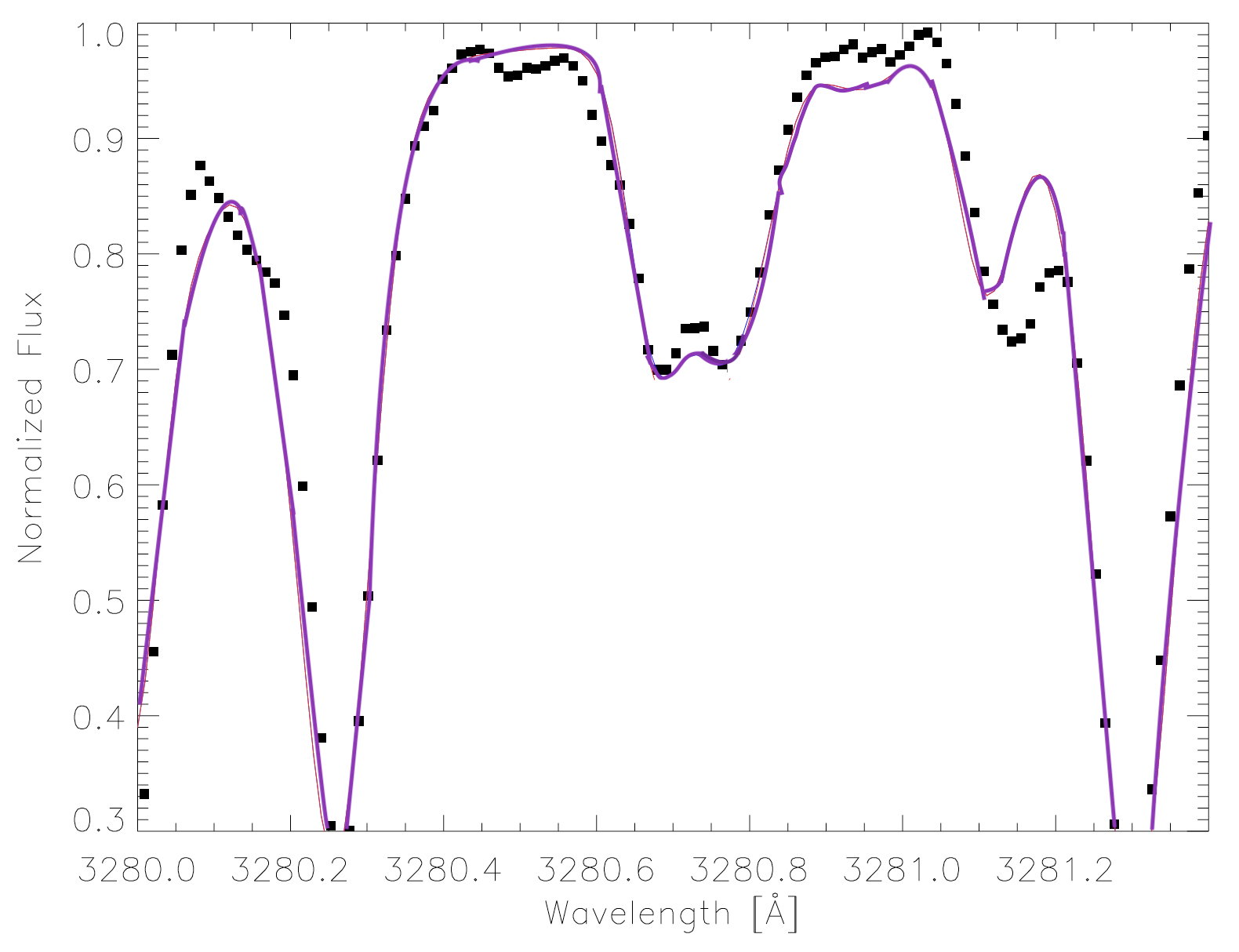}
 \includegraphics[width=0.96\textwidth]{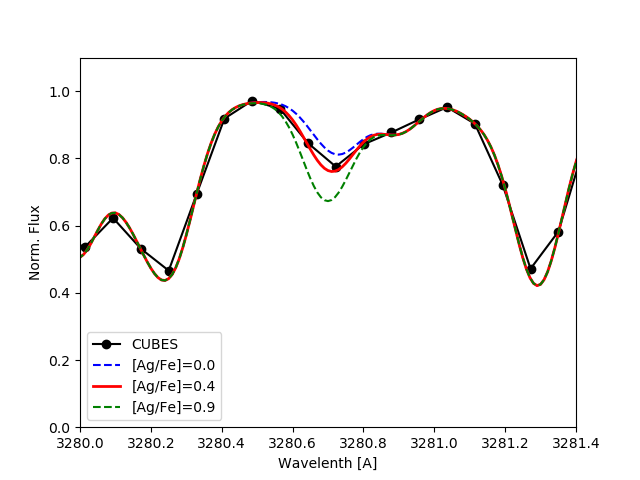}
\caption{Top: UVES spectrum of HD121004 (dwarf star) including spectrum synthesis of Ag and blending Zr and Fe lines (adapted from \cite{Hansen2012}). Bottom: Simulated CUBES spectra of a star like HD121004 with SNR $\sim$300 compared to three spectrum syntheses (see legend for details).}
\label{fig:Agspec}       
\end{figure}

The bluest range down to the atmospheric cutoff can only be observed with a handful of instruments mounted on large telescopes (e.g., UVES/VLT, Hires/Keck, HDS/Subaru, and down to $\sim3500$\,\AA\, MIKE/Magellan). Additionally, it is very challenging to design multi-object spectrographs that are efficient and have a high throughput in this blue region. This alone places limitations on the currently available sample sizes and how far away stars we can observe at the needed SNR to detect the often weak lines of heavy elements. A good example of this is silver, which shows its strongest transitions around 3300\,\AA\, (see Fig.~\ref{fig:Agspec}). Owing to the strong line blending at this wavelength, a high SNR is needed combined with a high resolution of the stellar spectra; these are crucial to derive abundances that are precise and accurate to within $~0.2$\,dex. In \cite{Hansen2012} we found that a SNR of 50 (100) per pixel would be needed to derive Ag abundances from giants (dwarfs), respectively, in UVES/VLT spectra (see Sect.~\ref{sec:cubes} below).
Accurate and precise abundances in larger samples ($>100$ stars) are needed to derive accurate stellar abundance patterns of many ($>10$) heavy elements as well as to probe possible abundance correlations in order to understand differences or similarities in their nuclear formation processes. Such correlations have amongst other things shown the clear need for two primary processes (main and weak r-process) to explain the production of on one hand Ag and Ru (weak r) and on the other hand Eu (main r) \cite{Hansen2012,Hansen2014a}.

\section{Heavy elements -- CUBES in context}\label{sec:cubes}

The UVES/VLT instrument with its broad wavelength coverage using a dichroic and high resolution has been used for many studies of heavy elements in metal-poor stars. An example of this is the study of silver (3280, 3382\,\AA\,) which shows its strongest transitions in the near-UV \cite{Hansen2012}. To study weak, blended lines, high resolution and high SNR are normally required, which typically leads to long exposure times. In Fig.~\ref{fig:Agspec} a UVES spectrum of the metal-poor dwarf star HD121004 is seen with a zoom in on an Ag line. To detect and accurately measure Ag abundances from these lines a SNR (per pixel) of 100 in dwarfs and 50 in giants would be needed with UVES. Owing to the lower gravity of the giants, these more evolved stars are typically targeted for heavy element studies since the giants' expanded atmospheres ease the heavy element line detections. This can lead to slightly biased abundances suffering from stellar evolutionary effects from mixing events. Depending on the magnitude of the star, this can result in very long integration times for giants but especially for dwarf stars. For HD121004 (Vmag=9) the integration with UVES of 1h resulted in a SNR $\sim 120$ around the Ag lines. For comparison, 1-hour integration with CUBES would yield a SNR of 100 in a 16-magnitude star, or for a dwarf like HD121004, the SNR would be reached within a few seconds to minutes. This opens new opportunities to study less evolved stars in larger numbers providing us with a cleaner trace and allowing us to probe the chemistry in more remote stars than currently feasible (at least within reasonable integration time). 

\section{Conclusion}
An instrument like CUBES will vastly reduce the cost (by reducing the integration time per target) of heavy element studies and hopefully finally provide us with observational probes that will help unveil the microphysics (e.g., neutron density or entropy in the environment) and the nature of some of the poorest known heavy element formation sites and reactions. This is particularly important in the era of large facilities like FAIR (Facility for Antiproton and Ion Research) which will provide new ways to experimentally explore a large region of the terra incognita of unknown n-rich reactions. The Extremely Large Telescope (ELT) will provide an unprecedented light-collecting power, however, the design of its instruments does not include the near-UV-blue wavelength range, leaving us blind in this range. As shown here, the blue spectral range is extremely important to better understand a number of interesting science cases including heavy elements - e.g., CEMP or r-II stars to mention a few. A high-resolution spectrograph providing high-SNR spectra is much needed to further our understanding in these directions, and accurate and precise abundances are key, showing the need for careful treatment of blends. Here a link to UVES or X-Shooter could help providing information on the blending component from molecular bands or clean lines from the redder wavelengths that would allow for a better deblending of heavy element blue lines which do not have detectable counter parts in the red. Improved stellar abundances will through element ratios and their correlations, or more complete abundance patterns, be able to provide traces of the first stars all the way up to stars formed after the Sun. Such observations will in turn help us map the nuclear reaction processes as well as help solve unknowns in large scale chemical enrichment models.  The interdisciplinary approach is much needed to make advances in the field of nuclear astrophysics and obtain a better understanding of the heavy elements and neutron-capture processes.

%
%

\begin{acknowledgements}
CJH acknowledges support from the Hessian collaborative research cluster ELEMENTS 
and the Deutsche Forschungsgemeinschaft (DFG, German Research Foundation) - Project-ID 279384907 - SFB 1245. CJH
would like to thank the Max Planck Society and H. Ernandes for spectra.
\end{acknowledgements}

%
 \section*{Conflict of interest}
 The authors declare that they have no conflict of interest and comply with Ethical Standards.

\bibliographystyle{spphys}       
\bibliography{Heavy_elements_Hansen}   

\begin{thebibliography}{10}
\providecommand{\url}[1]{{#1}}
\providecommand{\urlprefix}{URL }
\expandafter\ifx\csname urlstyle\endcsname\relax
  \providecommand{\doi}[1]{DOI \discretionary{}{}{}#1}\else
  \providecommand{\doi}{DOI \discretionary{}{}{}\begingroup
  \urlstyle{rm}\Url}\fi

\bibitem{Kobayashi2006}
C.~{Kobayashi}, H.~{Umeda}, K.~{Nomoto}, N.~{Tominaga}, T.~{Ohkubo}, \apj
  \textbf{653}, 1145 (2006).
\newblock \doi{10.1086/508914}

\bibitem{Francois2007}
P.~{Fran{\c c}ois}, E.~{Depagne}, V.~{Hill}, M.~{Spite}, F.~{Spite}, B.~{Plez},
  T.C. {Beers}, J.~{Andersen}, G.~{James}, B.~{Barbuy}, R.~{Cayrel},
  P.~{Bonifacio}, P.~{Molaro}, B.~{Nordstr{\"o}m}, F.~{Primas}, \aap
  \textbf{476}, 935 (2007).
\newblock \doi{10.1051/0004-6361:20077706}

\bibitem{Cescutti2008}
G.~{Cescutti}, \aap \textbf{481}, 691 (2008).
\newblock \doi{10.1051/0004-6361:20078571}

\bibitem{Hansen2012}
C.J. {Hansen}, F.~{Primas}, H.~{Hartman}, K.L. {Kratz}, S.~{Wanajo},
  B.~{Leibundgut}, K.~{Farouqi}, O.~{Hallmann}, N.~{Christlieb}, H.~{Nilsson},
  \aap \textbf{545}, A31 (2012).
\newblock \doi{10.1051/0004-6361/201118643}

\bibitem{Roederer2013}
I.U. {Roederer}, \aj \textbf{145}, 26 (2013).
\newblock \doi{10.1088/0004-6256/145/1/26}

\bibitem{Hansen2014b}
C.J. {Hansen}, F.~{Montes}, A.~{Arcones}, \apj \textbf{797}, 123 (2014).
\newblock \doi{10.1088/0004-637X/797/2/123}

\bibitem{Pruet2006}
J.~Pruet, R.D. Hoffman, S.E. Woosley, H.T. Janka, R.~Buras, \apj \textbf{644},
  1028 (2006)

\bibitem{Froehlich2006}
C.~{Fr{\"o}hlich}, G.~{Mart{\'{\i}}nez-Pinedo}, M.~{Liebend{\"o}rfer}, F.K.
  {Thielemann}, E.~{Bravo}, W.R. {Hix}, K.~{Langanke}, N.T. {Zinner}, Physical
  Review Letters \textbf{96}(14), 142502 (2006).
\newblock \doi{10.1103/PhysRevLett.96.142502}

\bibitem{Roederer2014}
I.U. {Roederer}, G.W. {Preston}, I.B. {Thompson}, S.A. {Shectman}, C.~{Sneden},
  G.S. {Burley}, D.D. {Kelson}, \aj \textbf{147}, 136 (2014).
\newblock \doi{10.1088/0004-6256/147/6/136}

\bibitem{Aoki2013}
W.~{Aoki}, T.~{Suda}, R.N. {Boyd}, T.~{Kajino}, M.A. {Famiano}, \apjl
  \textbf{766}(1), L13 (2013).
\newblock \doi{10.1088/2041-8205/766/1/L13}

\bibitem{Hansen2019}
C.J. {Hansen}, T.T. {Hansen}, A.~{Koch}, T.C. {Beers}, B.~{Nordstr{\"o}m}, V.M.
  {Placco}, J.~{Andersen}, \aap \textbf{623}, A128 (2019).
\newblock \doi{10.1051/0004-6361/201834601}

\bibitem{Koch2008}
A.~{Koch}, A.~{McWilliam}, E.K. {Grebel}, D.B. {Zucker}, V.~{Belokurov}, \apjl
  \textbf{688}, L13 (2008).
\newblock \doi{10.1086/595001}

\bibitem{Koch2013}
A.~{Koch}, S.~{Feltzing}, D.~{Ad{\'e}n}, F.~{Matteucci}, \aap \textbf{554}, A5 (2013).
\newblock \doi{10.1051/0004-6361/201220742}

\bibitem{Mashonkina2017b}
L.~{Mashonkina}, P.~{Jablonka}, T.~{Sitnova}, Y.~{Pakhomov}, P.~{North}, \aap
  \textbf{608}, A89 (2017).
\newblock \doi{10.1051/0004-6361/201731582}

\bibitem{Hansen2018}
C.J. {Hansen}, M.~{El-Souri}, L.~{Monaco}, S.~{Villanova}, P.~{Bonifacio},
  E.~{Caffau}, L.~{Sbordone}, \apj \textbf{855}, 83 (2018).
\newblock \doi{10.3847/1538-4357/aa978f}

\bibitem{Hill2002}
V.~{Hill}, B.~{Plez}, R.~{Cayrel}, T.C. {Beers}, B.~{Nordstr{\"o}m},
  J.~{Andersen}, M.~{Spite}, F.~{Spite}, B.~{Barbuy}, P.~{Bonifacio},
  E.~{Depagne}, P.~{Fran{\c c}ois}, F.~{Primas}, \aap \textbf{387}, 560 (2002).
\newblock \doi{10.1051/0004-6361:20020434}

\bibitem{Sneden2003}
C.~{Sneden}, J.J. {Cowan}, J.E. {Lawler}, I.I. {Ivans}, S.~{Burles}, T.C.
  {Beers}, F.~{Primas}, V.~{Hill}, J.W. {Truran}, G.M. {Fuller}, B.~{Pfeiffer},
  K.L. {Kratz}, \apj \textbf{591}, 936 (2003).
\newblock \doi{10.1086/375491}

\bibitem{Beers2005}
T.C. {Beers}, N.~{Christlieb}, \araa \textbf{43}, 531 (2005).
\newblock \doi{10.1146/annurev.astro.42.053102.134057}

\bibitem{Holmbeck2020}
E.~M.~{Holmbeck}, T.~T.~{Hansen}, T.~C.~{Beers}, V.~C.~{Placco}, D.~D.~{Whitten}, K.~C.~{Rasmussen}, I.~U.~{Roederer}, R.~{Ezzeddine}, C.~M.~{Sakari}, A.~{Frebel}, M.~R.~{Drout}, J.~D.~{Simon}, I.~B.~{Thompson}, J.~{Bland-Hawthorn}, B.~K.~{Gibson}, E.~K.~{Grebel}, G.~{Kordopatis}, A.~{Kunder}, J.~{Mel{\'e}ndez}, J.~F.~{Navarro}, W.~A.~{Reid},G.~{Seabroke}, M.~{Steinmetz}, F.~{Watson}, R.~F.A{Wyse}, \apjs \textbf{249} 30 (2020)
\newblock \doi{10.3847/1538-4365/ab9c19}

\bibitem{Schatz2002}
H.~{Schatz}, R.~{Toenjes}, B.~{Pfeiffer}, T.C. {Beers}, J.J. {Cowan},
  V.~{Hill}, K.L. {Kratz}, \apj \textbf{579}, 626 (2002).
\newblock \doi{10.1086/342939}

\bibitem{Cowan2002}
J.J. {Cowan}, C.~{Sneden}, S.~{Burles}, I.I. {Ivans}, T.C. {Beers}, J.W.
  {Truran}, J.E. {Lawler}, F.~{Primas}, G.M. {Fuller}, B.~{Pfeiffer}, K.L.
  {Kratz}, \apj \textbf{572}, 861 (2002).
\newblock \doi{10.1086/340347}

\bibitem{Barbuy2011}
B.~{Barbuy}, M.~{Spite}, V.~{Hill}, F.~{Primas}, B.~{Plez}, R.~{Cayrel},
  F.~{Spite}, S.~{Wanajo}, C.~{Siqueira Mello}, J.~{Andersen},
  B.~{Nordstr{\"o}m}, T.C. {Beers}, P.~{Bonifacio}, P.~{Fran{\c{c}}ois},
  P.~{Molaro}, \aap \textbf{534}, A60 (2011).
\newblock \doi{10.1051/0004-6361/201117450}

\bibitem{Hansen2018RPA}
T.T. {Hansen}, E.M. {Holmbeck}, T.C. {Beers}, V.M. {Placco}, I.U. {Roederer},
  A.~{Frebel}, C.M. {Sakari}, J.D. {Simon}, I.B. {Thompson}, \apj
  \textbf{858}(2), 92 (2018).
\newblock \doi{10.3847/1538-4357/aabacc}

\bibitem{Sakari2018}
C.M. {Sakari}, V.M. {Placco}, E.M. {Farrell}, I.U. {Roederer},
  G.~{Wallerstein}, T.C. {Beers}, R.~{Ezzeddine}, A.~{Frebel}, T.~{Hansen},
  E.M. {Holmbeck}, C.~{Sneden}, J.J. {Cowan}, K.A. {Venn}, C.E. {Davis},
  G.~{Matijevi{\v{c}}}, R.F.G. {Wyse}, J.~{Bland-Hawthorn}, C.~{Chiappini},
  K.C. {Freeman}, B.K. {Gibson}, E.K. {Grebel}, A.~{Helmi}, G.~{Kordopatis},
  A.~{Kunder}, J.~{Navarro}, W.~{Reid}, G.~{Seabroke}, M.~{Steinmetz},
  F.~{Watson}, \apj \textbf{868}(2), 110 (2018).
\newblock \doi{10.3847/1538-4357/aae9df}

\bibitem{Honda2004}
S.~{Honda}, W.~{Aoki}, T.~{Kajino}, H.~{Ando}, T.C. {Beers}, H.~{Izumiura},
  K.~{Sadakane}, M.~{Takada-Hidai}, \apj \textbf{607}, 474 (2004).
\newblock \doi{10.1086/383406}

\bibitem{Honda2007}
S.~{Honda}, W.~{Aoki}, Y.~{Ishimaru}, S.~{Wanajo}, \apj \textbf{666}, 1189
  (2007).
\newblock \doi{10.1086/520034}

\bibitem{Peterson2013}
R.C. {Peterson}, \apjl \textbf{768}(1), L13 (2013).
\newblock \doi{10.1088/2041-8205/768/1/L13}

\bibitem{Hansen2014a}
C.J. {Hansen}, A.C. {Andersen}, N.~{Christlieb}, \aap \textbf{568}, A47 (2014).
\newblock \doi{10.1051/0004-6361/201423535}

\bibitem{Peterson2020}
 R.~C.~{Peterson}, B.~{Barbuy}, M.~{Spite}, \aap \textbf{638} A64 (2020)
 \newblock \doi{10.1051/0004-6361/202037689}


\bibitem{Merrill1952}
P.W. {Merrill}, \apj \textbf{116}, 21 (1952).
\newblock \doi{10.1086/145589}

\bibitem{VanEck1999}
S.~{Van Eck}, A.~{Jorissen}, \aap \textbf{345}, 127 (1999)

\bibitem{Neyskens2015}
P.~{Neyskens}, S.~{van Eck}, A.~{Jorissen}, S.~{Goriely}, L.~{Siess},
  B.~{Plez}, \nat \textbf{517}(7533), 174 (2015).
\newblock \doi{10.1038/nature14050}

\bibitem{Busso1999}
M.~{Busso}, R.~{Gallino}, G.J. {Wasserburg}, \araa \textbf{37}, 239 (1999).
\newblock \doi{10.1146/annurev.astro.37.1.239}

\bibitem{Kappeler2011}
F.~K{\"a}ppeler, R.~Gallino, S.~Bisterzo, W.~Aoki, Review of Modern Physics
  \textbf{83}(1), 157 (2011)

\bibitem{Maeder2003}
A.~{Maeder}, G.~{Meynet}, \aap \textbf{411}, 543 (2003).
\newblock \doi{10.1051/0004-6361:20031491}

\bibitem{Meynet2006}
G.~{Meynet}, S.~{Ekstr{\"o}m}, A.~{Maeder}, \aap \textbf{447}, 623 (2006).
\newblock \doi{10.1051/0004-6361:20053070}

\bibitem{Hirschi2007}
R.~{Hirschi}, \aap \textbf{461}, 571 (2007).
\newblock \doi{10.1051/0004-6361:20065356}

\bibitem{Frischknecht2012}
U.~{Frischknecht}, R.~{Hirschi}, F.K. {Thielemann}, \aap \textbf{538}, L2
  (2012).
\newblock \doi{10.1051/0004-6361/201117794}

\bibitem{Pignatari2013}
M.~{Pignatari}, R.~{Hirschi}, M.~{Wiescher}, R.~{Gallino}, M.~{Bennett},
  M.~{Beard}, C.~{Fryer}, F.~{Herwig}, G.~{Rockefeller}, F.X. {Timmes}, \apj
  \textbf{762}, 31 (2013).
\newblock \doi{10.1088/0004-637X/762/1/31}

\bibitem{VanEck2001}
S.~{Van Eck}, S.~{Goriely}, A.~{Jorissen}, B.~{Plez}, \nat \textbf{412}(6849),
  793 (2001).
\newblock \doi{10.1038/35090514}

\bibitem{CowanRose1977}
J.J. {Cowan}, W.K. {Rose}, \apj \textbf{212}, 149 (1977).
\newblock \doi{10.1086/155030}

\bibitem{Herwig2011}
F.~{Herwig}, M.~{Pignatari}, P.R. {Woodward}, D.H. {Porter}, G.~{Rockefeller},
  C.L. {Fryer}, M.~{Bennett}, R.~{Hirschi}, \apj \textbf{727}(2), 89 (2011).
\newblock \doi{10.1088/0004-637X/727/2/89}

\bibitem{Hampel2016}
M.~{Hampel}, R.J. {Stancliffe}, M.~{Lugaro}, B.S. {Meyer}, \apj \textbf{831},
  171 (2016).
\newblock \doi{10.3847/0004-637X/831/2/171}

\bibitem{Denissenkov2019}
P.A. {Denissenkov}, F.~{Herwig}, P.~{Woodward}, R.~{Andrassy}, M.~{Pignatari},
  S.~{Jones}, \mnras \textbf{488}(3), 4258 (2019).
\newblock \doi{10.1093/mnras/stz1921}

\bibitem{Roederer2016}
I.U. {Roederer}, A.I. {Karakas}, M.~{Pignatari}, F.~{Herwig}, \apj
  \textbf{821}(1), 37 (2016).
\newblock \doi{10.3847/0004-637X/821/1/37}

\bibitem{Mishenina2015}
T.~{Mishenina}, M.~{Pignatari}, G.~{Carraro}, V.~{Kovtyukh}, L.~{Monaco},
  S.~{Korotin}, E.~{Shereta}, I.~{Yegorova}, F.~{Herwig}, \mnras
  \textbf{446}(4), 3651 (2015).
\newblock \doi{10.1093/mnras/stu2337}

\bibitem{Koch2019i}
A.~{Koch}, M.~{Reichert}, C.J. {Hansen}, M.~{Hampel}, R.J. {Stancliffe},
  A.~{Karakas}, A.~{Arcones}, \aap \textbf{622}, A159 (2019).
\newblock \doi{10.1051/0004-6361/201834241}

\bibitem{Spite2005}
M.~{Spite}, R.~{Cayrel}, B.~{Plez}, V.~{Hill}, F.~{Spite}, E.~{Depagne},
  P.~{Fran{\c c}ois}, P.~{Bonifacio}, B.~{Barbuy}, T.~{Beers}, J.~{Andersen},
  P.~{Molaro}, B.~{Nordstr{\"o}m}, F.~{Primas}, \aap \textbf{430}, 655 (2005).
\newblock \doi{10.1051/0004-6361:20041274}

\bibitem{Lee2013}
Y.S. {Lee}, T.C. {Beers}, T.~{Masseron}, B.~{Plez}, C.M. {Rockosi},
  J.~{Sobeck}, B.~{Yanny}, S.~{Lucatello}, T.~{Sivarani}, V.M. {Placco},
  D.~{Carollo}, \aj \textbf{146}, 132 (2013).
\newblock \doi{10.1088/0004-6256/146/5/132}

\bibitem{Caffau2011}
E.~{Caffau}, P.~{Bonifacio}, P.~{Fran{\c c}ois}, L.~{Sbordone}, L.~{Monaco},
  M.~{Spite}, F.~{Spite}, H.G. {Ludwig}, R.~{Cayrel}, S.~{Zaggia}, F.~{Hammer},
  S.~{Randich}, P.~{Molaro}, V.~{Hill}, \nat \textbf{477}, 67 (2011).
\newblock \doi{10.1038/nature10377}

\bibitem{Masseron2010}
T.~{Masseron}, J.A. {Johnson}, B.~{Plez}, S.~{van Eck}, F.~{Primas},
  S.~{Goriely}, A.~{Jorissen}, \aap \textbf{509}, A93 (2010).
\newblock \doi{10.1051/0004-6361/200911744}

\bibitem{Winteler2012}
C.~{Winteler}, R.~{K{\"a}ppeli}, A.~{Perego}, A.~{Arcones}, N.~{Vasset},
  N.~{Nishimura}, M.~{Liebend{\"o}rfer}, F.K. {Thielemann}, \apjl \textbf{750},
  L22 (2012).
\newblock \doi{10.1088/2041-8205/750/1/L22}

\bibitem{Lugaro2012}
M.~{Lugaro}, A.I. {Karakas}, R.J. {Stancliffe}, C.~{Rijs}, \apj \textbf{747}, 2
  (2012).
\newblock \doi{10.1088/0004-637X/747/1/2}

\bibitem{Karakas2014}
A.I. {Karakas}, J.C. {Lattanzio}, \pasa \textbf{31}, e030 (2014).
\newblock \doi{10.1017/pasa.2014.21}

\bibitem{Watson2019}
D.~{Watson}, C.J. {Hansen}, J.~{Selsing}, A.~{Koch}, D.B. {Malesani}, A.C.
  {Andersen}, J.P.U. {Fynbo}, A.~{Arcones}, A.~{Bauswein}, S.~{Covino},
  A.~{Grado}, K.E. {Heintz}, L.~{Hunt}, C.~{Kouveliotou}, G.~{Leloudas}, A.J.
  {Levan}, P.~{Mazzali}, E.~{Pian}, \nat \textbf{574}(7779), 497 (2019).
\newblock \doi{10.1038/s41586-019-1676-3}

\bibitem{Korotin2015}
S.A. {Korotin}, S.M. {Andrievsky}, C.J. {Hansen}, E.~{Caffau}, P.~{Bonifacio},
  M.~{Spite}, F.~{Spite}, P.~{Fran{\c c}ois}, \aap \textbf{581}, A70 (2015).
\newblock \doi{10.1051/0004-6361/201526558}

\bibitem{Andrievsky2009}
S.M. {Andrievsky}, M.~{Spite}, S.A. {Korotin}, F.~{Spite}, P.~{Fran{\c c}ois},
  P.~{Bonifacio}, R.~{Cayrel}, V.~{Hill}, \aap \textbf{494}, 1083 (2009).
\newblock \doi{10.1051/0004-6361:200810894}

\bibitem{Gallagher2020}
A.J. {Gallagher}, M.~{Bergemann}, R.~{Collet}, B.~{Plez}, J.~{Leenaarts},
  M.~{Carlsson}, S.A. {Yakovleva}, A.K. {Belyaev}, \aap \textbf{634}, A55
  (2020).
\newblock \doi{10.1051/0004-6361/201936104}

\bibitem{Aguado2021}
 D.~S.~Aguado, V.~Belokurov,  G.~C.~Myeong, N.~W.~{Evans},  C.~{Kobayashi}, L.~{Sbordone},  J.~{Chanam{\'e}}, C.~{Navarrete}, S.~E.~{Koposov}, \apjl \textbf{908}, L8 (2021). \newblock \doi{10.3847/2041-8213/abdbb8}

\bibitem{Ji2016}
A.P. {Ji}, A.~{Frebel}, A.~{Chiti}, J.D. {Simon}, \nat \textbf{531}(7596), 610
  (2016).
\newblock \doi{10.1038/nature17425}

\bibitem{Reichert2021}
M.~{Reichert}, C.J. {Hansen}, A.~{Arcones}, \apj \textbf{912}(2), 157 (2021).
\newblock \doi{10.3847/1538-4357/abefd8}

\bibitem{Kobayashi2020}
C.~{Kobayashi}, A.I. {Karakas}, M.~{Lugaro}, \apj \textbf{900}(2), 179 (2020).
\newblock \doi{10.3847/1538-4357/abae65}

\bibitem{Cote2019}
B.~{C{\^o}t{\'e}}, M.~{Eichler}, A.~{Arcones}, C.J. {Hansen}, P.~{Simonetti},
  A.~{Frebel}, C.L. {Fryer}, M.~{Pignatari}, M.~{Reichert}, K.~{Belczynski},
  F.~{Matteucci}, \apj \textbf{875}(2), 106 (2019).
\newblock \doi{10.3847/1538-4357/ab10db}

\bibitem{Spina2018}
L.~{Spina}, J.~{Mel{\'e}ndez}, A.I. {Karakas}, L.~{dos Santos}, M.~{Bedell},
  M.~{Asplund}, I.~{Ram{\'\i}rez}, D.~{Yong}, A.~{Alves-Brito}, J.L. {Bean},
  S.~{Dreizler}, \mnras \textbf{474}(2), 2580 (2018).
\newblock \doi{10.1093/mnras/stx2938}

\bibitem{Asplund2021}
M.~{Asplund}, A.M. {Amarsi}, N.~{Grevesse}, \aap \textbf{653}, A141 (2021).
\newblock \doi{10.1051/0004-6361/202140445}

\bibitem{Caffau2015O}
E.~{Caffau}, H.G. {Ludwig}, M.~{Steffen}, W.~{Livingston}, P.~{Bonifacio}, J.M.
  {Malherbe}, H.P. {Doerr}, W.~{Schmidt}, \aap \textbf{579}, A88 (2015).
\newblock \doi{10.1051/0004-6361/201526331}

\bibitem{Bergemann2021}
M.~{Bergemann}, R.~{Hoppe}, E.~{Semenova}, M.~{Carlsson}, S.A. {Yakovleva},
  Y.V. {Voronov}, M.~{Bautista}, A.~{Nemer}, A.K. {Belyaev}, J.~{Leenaarts},
  L.~{Mashonkina}, A.~{Reiners}, M.~{Ellwarth}, arXiv e-prints arXiv:2109.01143
  (2021)

\bibitem{Bergemann2012}
M.~{Bergemann}, C.J. {Hansen}, M.~{Bautista}, G.~{Ruchti}, \aap \textbf{546},
  A90 (2012).
\newblock \doi{10.1051/0004-6361/201219406}

\bibitem{Hansen2013}
C.J. {Hansen}, M.~{Bergemann}, G.~{Cescutti}, P.~{Fran{\c c}ois}, A.~{Arcones},
  A.I. {Karakas}, K.~{Lind}, C.~{Chiappini}, \aap \textbf{551}, A57 (2013).
\newblock \doi{10.1051/0004-6361/201220584}

\bibitem{Hansen2020}
C.J. {Hansen}, A.~{Koch}, L.~{Mashonkina}, M.~{Magg}, M.~{Bergemann},
  T.~{Sitnova}, A.J. {Gallagher}, I.~{Ilyin}, E.~{Caffau}, H.W. {Zhang}, K.G.
  {Strassmeier}, R.S. {Klessen}, \aap \textbf{643}, A49 (2020).
\newblock \doi{10.1051/0004-6361/202038805}

\bibitem{Hansen2015}
C.J. {Hansen}, H.G. {Ludwig}, W.~{Seifert}, A.~{Koch}, W.~{Xu}, E.~{Caffau},
  N.~{Christlieb}, A.J. {Korn}, K.~{Lind}, L.~{Sbordone}, G.~{Ruchti},
  S.~{Feltzing}, R.S. {de Jong}, S.~{Barden}, Astronomische Nachrichten
  \textbf{336}, 665 (2015).
\newblock \doi{10.1002/asna.201512206}

\bibitem{Caffau2013}
E.~{Caffau}, A.~{Koch}, L.~{Sbordone}, P.~{Sartoretti}, C.J. {Hansen},
  F.~{Royer}, N.~{Leclerc}, P.~{Bonifacio}, N.~{Christlieb}, H.G. {Ludwig},
  E.K. {Grebel}, R.S. {de Jong}, C.~{Chiappini}, J.~{Walcher}, S.~{Mignot},
  S.~{Feltzing}, M.~{Cohen}, I.~{Minchev}, A.~{Helmi}, T.~{Piffl},
  E.~{Depagne}, O.~{Schnurr}, Astronomische Nachrichten \textbf{334}, 197
  (2013).
\newblock \doi{10.1002/asna.201211814}

\bibitem{Jong2014}
R.S. {de Jong}, S.~{Barden}, O.~{Bellido-Tirado}, J.~{Brynnel}, C.~{Chiappini},
  {\'E}.~{Depagne}, R.~{Haynes}, D.~{Johl}, D.P. {Phillips}, O.~{Schnurr}, A.D.
  {Schwope}, J.~{Walcher}, S.M. {Bauer}, G.~{Cescutti}, M.R.L. {Cioni},
  F.~{Dionies}, H.~{Enke}, D.M. {Haynes}, A.~{Kelz}, F.S. {Kitaura},
  G.~{Lamer}, I.~{Minchev}, V.~{M{\"u}ller}, S.E. {Nuza}, J.C. {Olaya},
  T.~{Piffl}, E.~{Popow}, A.~{Saviauk}, M.~{Steinmetz}, U.~{Ural},
  M.~{Valentini}, R.~{Winkler}, L.~{Wisotzki}, W.R. {Ansorge}, M.~{Banerji},
  E.~{Gonzalez Solares}, M.~{Irwin}, R.C. {Kennicutt}, D.M.P. {King},
  R.~{McMahon}, S.~{Koposov}, I.R. {Parry}, X.~{Sun}, N.A. {Walton},
  G.~{Finger}, O.~{Iwert}, M.~{Krumpe}, J.L. {Lizon}, V.~{Mainieri}, J.P.
  {Amans}, P.~{Bonifacio}, M.~{Cohen}, P.~{Fran{\c c}ois}, P.~{Jagourel}, S.B.
  {Mignot}, F.~{Royer}, P.~{Sartoretti}, R.~{Bender}, H.J. {Hess},
  F.~{Lang-Bardl}, B.~{Muschielok}, J.~{Schlichter}, H.~{B{\"o}hringer},
  T.~{Boller}, A.~{Bongiorno}, M.~{Brusa}, T.~{Dwelly}, A.~{Merloni},
  K.~{Nandra}, M.~{Salvato}, J.H. {Pragt}, R.~{Navarro}, G.~{Gerlofsma},
  R.~{Roelfsema}, G.B. {Dalton}, K.F. {Middleton}, I.A. {Tosh}, C.~{Boeche},
  E.~{Caffau}, N.~{Christlieb}, E.K. {Grebel}, C.J. {Hansen}, A.~{Koch}, H.G.
  {Ludwig}, H.~{Mandel}, A.~{Quirrenbach}, L.~{Sbordone}, W.~{Seifert},
  G.~{Thimm}, A.~{Helmi}, S.C. {trager}, T.~{Bensby}, S.~{Feltzing},
  G.~{Ruchti}, B.~{Edvardsson}, A.~{Korn}, K.~{Lind}, W.~{Boland},
  M.~{Colless}, G.~{Frost}, J.~{Gilbert}, P.~{Gillingham}, J.~{Lawrence},
  N.~{Legg}, W.~{Saunders}, A.~{Sheinis}, S.~{Driver}, A.~{Robotham},
  R.~{Bacon}, P.~{Caillier}, J.~{Kosmalski}, F.~{Laurent}, J.~{Richard}, in
  \emph{Society of Photo-Optical Instrumentation Engineers (SPIE) Conference
  Series}, \emph{Society of Photo-Optical Instrumentation Engineers (SPIE)
  Conference Series}, vol. 9147 (2014), \emph{Society of Photo-Optical
  Instrumentation Engineers (SPIE) Conference Series}, vol. 9147, p.~0.
\newblock \doi{10.1117/12.2055826}

\end{thebibliography}

%
%


\end{document}